\documentclass[10pt,letterpaper]{article}
\usepackage[OME]{express}
\usepackage{graphicx}
\usepackage{csquotes}

\begin{document}

\title{Excitation of nanowire surface plasmons by silicon vacancy centers in nanodiamonds}

\author{Shailesh Kumar,\authormark{1,*} Valery A. Davydov, \authormark{2} Viatcheslav N. Agafonov, \authormark{3} and Sergey I. Bozhevolnyi\authormark{1}}
\address{\authormark{1}Centre for Nano Optics, University of Southern Denmark, Campusvej 55, Odense M, DK-5230, Denmark\\
\authormark{2}L.F. Vereshchagin Institute for High Pressure Physics, Russian Academy of Sciences, Troitsk, Moscow, 142190, Russia\\
\authormark{3}GREMAN, UMR CNRS CEA 6157, Universit\'e F. Rabelais, 37200 Tours, France}

\email{\authormark{*}shku@mci.sdu.dk}


\date{\today}

\begin*{abstract}
  Silicon vacancy (SiV) centers in diamonds have emerged as a very promising candidate for quantum emitter due to their narrow emission line resulting in their indistinguishability. While many different quantum emitters have already been used for excitation of various propagating plasmonic modes, the corresponding exploitation of SiV centers remained so far uncharted territory. Here, we report on excitation of surface plasmon modes supported by silver nanowires using SiV centers in nanodiamonds. The coupling of SiV center fluorescence to surface plasmons is observed, when a nanodiamond situated close to a nanowire is illuminated by the pump, as radiated emission from the distal nanowire end. The effect of coupling is also seen as a change in the SiV center lifetime. Finally, we discuss possible avenues for strengthening the SiV center coupling to surface plasmon modes.
\end*{abstract}

%%%%%%%%%%%%%%%%%%%%%%% References %%%%%%%%%%%%%%%%%%%%%%%%%

%%%%%%%%%%%%%%%%%%%%%%%%%%  body  %%%%%%%%%%%%%%%%%%%%%%%%%%
\section{Introduction}
Quantum emitter coupled to waveguides can be useful for devising an efficient source of single photons, entanglement of quantum emitters and eventually for quantum communication and quantum computing~\cite{Brien1567, QuantumInternet, 2006Chang}. There are many possible candidates for quantum emitters, such as, atoms - which have the advantage of being identical to each other, have a disadvantage of the requirement of trapping them~\cite{QuantumInternet}. Molecules - can be put very precisely in required places, but they tend to bleach~\cite{2009Nphotonbowtie}. Situation with quantum dots is similar, although quantum dot technology has been improving and research towards making quantum dots which do not blink or bleach is ongoing~\cite{2013NmaterQD}. Defect centers in diamond are quantum emitters, which have been shown to be stable. The most commonly used defect center in diamond, namely nitrogen vacancy (NV) centers, have been shown to be useful for certain quantum information related tasks~\cite{HongOuLukin, HongOuHanson, SpinPhotonEntNV, HansonEnt, Pfaff532, LHFHanson}. However, the emission from NV centers is broad due to phononic interaction, and only a small fraction of its emission is within the zero phonon line (ZPL)~\cite{Doherty20131}. This has been a limiting factor in extending the entanglement of NV centers beyond a couple of NV centers~\cite{HansonEnt}. One of the other defect centers in diamonds, silicon vacancy (SiV) center, has been shown to emit mostly into its ZPL~\cite{2013Elke, NcommFedor, PhysRevAppliedSiV2016, 2016Uwe}. It has been shown that differerent SiV centers emit indistinguishable photons~\cite{SiVHongOuM, NcommFedor}. These properties have recently been utilized in a nonlinear device operating at single photon level as well as the entanglement of two SiV centers~\cite{Sipahigil847}. This was realized by the placement of SiV center in a photonic crystal cavity, which formed an efficient interface between photons and SiV center. Coupling of quantum emitters to surface plasmon modes is promising from the point of view of advantageously exploiting extreme mode confinement ensured by various plasmonic waveguides. The coupled system is also proposed as a platform where nonlinear effects at single photon level can be studied and quantum emitters can be entangled~\cite{2007Chang, 2011Ent}. This will enable formation of a quantum network, and will eventually allow for quantum computation. Indeed, efficient coupling of quantum dots and NV centers in diamonds to nanowire~\cite{2007Akimov, 2009Kolesov, 2011Huck}, channel~\cite{2015Esteban} and wedge~\cite{2015DavidNorrisNL} surface plasmons has been demonstrated. At the same time, despite certain advantages of SiV centers (narrow emission spectrum and indistinguishability) their coupling to plasmonic waveguide modes insofar has not been realized~\cite{2016SSSPEreview}.

In this article, we demonstrate the coupling of SiV centers to strongly confined nanowire surface plasmons. For the coupled SiV centers, we observe a decrease in lifetime when compared to the lifetime of uncoupled SiV centers. The coupled system is further characterized by the measurement of spectrum and lifetime at the distal end. 

\section{Experiment and results}

To perform this experiment, we use chemically grown silver nanowires and nanodiamonds which have been grown with SiV centers. To grow the single crystalline silver nanowires, we have followed the method from Korte et. al.~\cite{2008Korte}. The silver wires thus obtained have a diameter ranging from 80~nm to 200~nm and lengths ranging from 5~$\mu$m to 20~$\mu$m. The nanodiamonds are grown at high temperature and high pressure, where Si containing compound is mixed in the process of growth of the nanodiamonds~\cite{Davydov2014}. This process can produce SiV center containing diamonds in sizes ranging from few nanometers to few micrometers.

\begin{figure}[htbp]
\centering\includegraphics{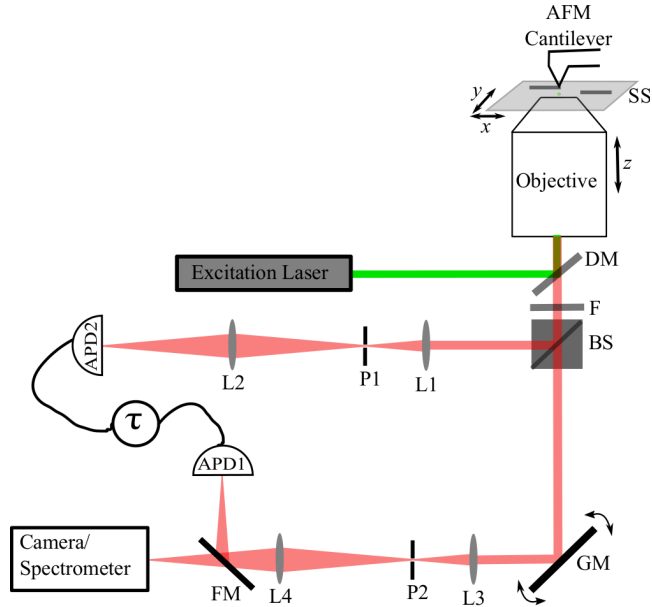}
\caption{Schematics of the experimental set-up.  SS:sample stage, DM: dichroic mirror, F: filters, BS: beam-splitter, L1, L2, L3, L4: lenses, P1, P2: pin-holes, GM: galvanometric mirrors, APD1 and APD2: avalanche photo diodes. APD1 and APD2 are connected to counting electronics which relates the detection timing of photons in the two channels ($\tau$). \label{figure1}}
\end{figure}

For characterization of the SiV centers and SiV-silver nanowire coupled system, we use an experimental set-up which combines an inverted confocal microscope and an atomic force microscope (AFM), a schematic of which is shown in Fig.~\ref{figure1}. A pulsed laser with a pulse rate variable between 2.5~MHz and 80~MHz with a pulsewidth of 50~ps, or a continuous wave (CW) excitation laser at a wavelength of 532~nm is used. The sample can be raster scanned in a plane (xy in Fig.~\ref{figure1}). We use an objective of numerical aperture (NA) 1.4 to focus the excitation laser onto the sample, and the fluorescence emission is collected from the same objective. A dichroic mirror is used to separate the excitation laser from the emission of quantum emitter. We use filters in detection path so that photons with wavelengths in the range 725~nm and 770~nm reach our detectors. A 50\slash 50 beam splitter is used to split the emission into two paths. In one path, the emission is spatially filtered with a pin-hole, and the pin-hole is imaged onto an avalanche photo-diode (APD). In another path, a galvanometric mirror is used to image the sample plane while exciting a quantum emitter. This feature helps in observation of coupling of quantum emitter emission to a waveguide. A spectrometer is used to obtain the emission spectrum from the quantum emitter as well as from the waveguide ends. The sample can also be scanned using an AFM to obtain the topography.

\begin{figure}[htbp]
\includegraphics{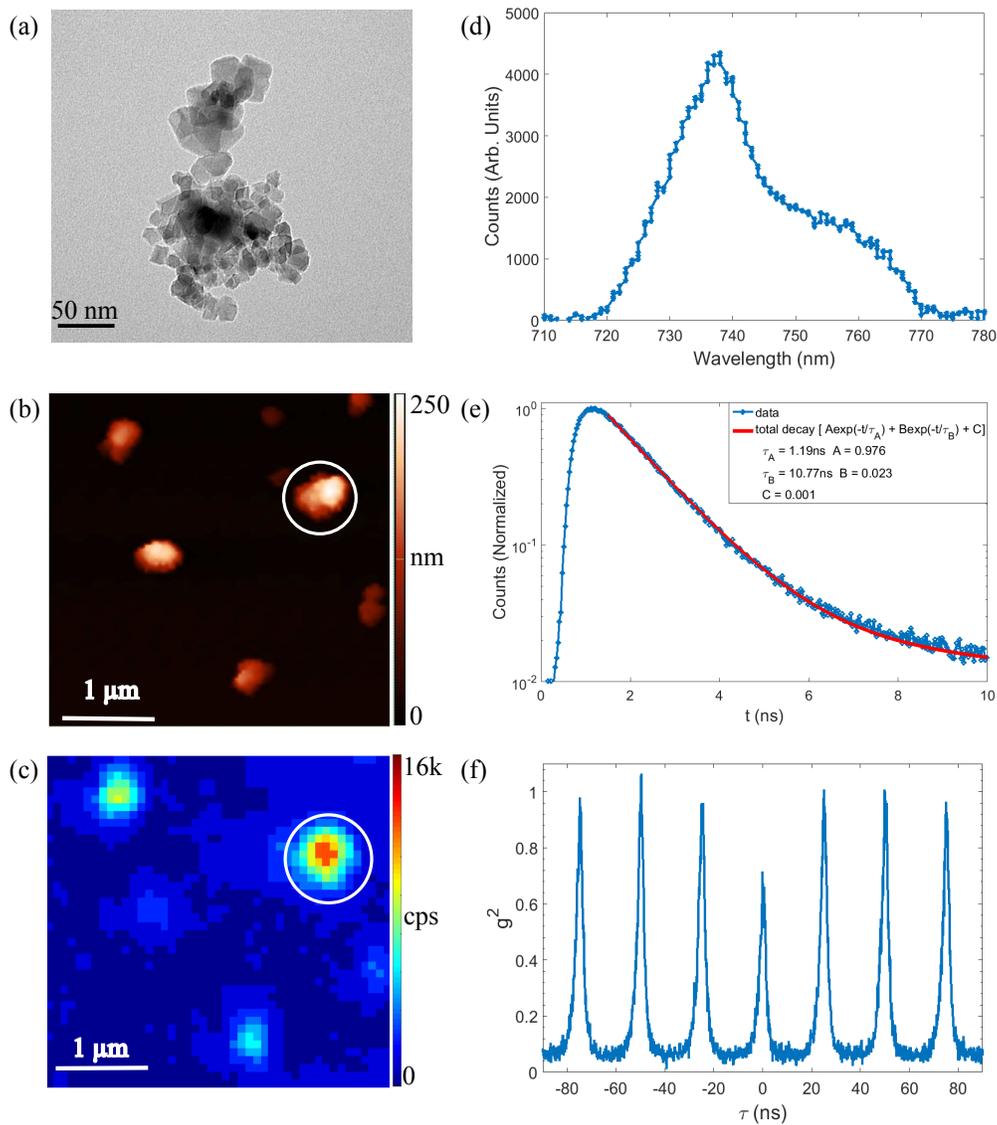}
\caption{(a) a TEM image and (b) an AFM image of nanodiamonds. (c) a fluorescence confocal microscope image of the same area, as in figure (b). (d) Spectrum obtained for the spot encircled in figure (b) and (c). (e) Measured data for the lifetime and a double exponential fit to the data. (f) Auto-correlation measured for the spot encircled in figure (b) and (c). \label{figure2}}
\end{figure}

First, we characterize the nanodiamonds that contain SiV centers. A Transmission electron microscope (TEM) image of nanodiamonds is shown in Fig.~\ref{figure2}(a),  and it can be observed that the nanodiamonds are small (sizes in the range of 10~nm and 50~nm), and they tend to cluster.  We further characterized the nanodiamonds, in the set-up described before, by spin coating them on a fused silica glass substrate.  An AFM image is shown in Fig.~\ref{figure2}(b), where the height of nanodiamond clusters are upto 300~nm. We then took a fluorescence scan image of the same area as presented in Fig.~\ref{figure2}(b) at an excitation power of 1mW of CW laser. The power used for all the fluorescence scans presented in this article is 1mW of CW laser. The fluorescence image, presented in Fig.~\ref{figure2}(c),  shows spots corresponding to the nanodiamond clusters. Subsequently, we characterized each of the spots by measuring  their spectrum, lifetime and second order auto-correlation. Spectrum for the spot encircled in Fig.~\ref{figure2}(c) is presented in Fig.~\ref{figure2}(d), which shows a typical spectrum for SiV center at room temperature~\cite{Beha2012,2014Naturenano,2010Elke}. Spectra measured for five different nanodiamonds containing SiV centers are presented in Appendix, where similarities and differences between different SiV center spectra can be seen. Next, lifetime for the encircled spot in Fig.~\ref{figure2}(c) is measured and tail-fitted to a double exponential decay curve. We fit the data with two exponentials, which is the minimum number of exponentials that fit the measured data well. The proportion of the low lifetime (1.19~ns) component is around 0.98. This is a typical lifetime obtained for SiV center in bulk diamond at room temperature~\cite{NcommFedor}. We observe similar lifetimes for SiV centers in our nanodiamonds, and the lifetime averaged over 10 such spots is 1.24~ns. Longer lifetime component (10.77~ns) with small amplitude (0.02) also shows up, which can be due to the presence of other defect centers in diamond (such as nitrogen vacancy center, which have lifetimes in this range). The SiV center spectrum and lifetime clearly suggest that SiV centers are predominantly present in the nanodiamonds. Furthermore, the auto-correlation for the encircled spot in  Fig.~\ref{figure2}(c) was subsequently measured, which we present in Fig.~\ref{figure2}(f). From the auto-correlation measurement, a $g^2(0) = 0.71$ can be observed, from which we infer that the number of SiV centers for this spot is 3. For other spots, in Fig.~\ref{figure2}(c), we did not observe any dip in the auto-correlation measurements. This is due to higher number of emitters in the spots, which may not be optimally aligned with respect to the excitation polarization. 

\begin{figure}[htbp]
\includegraphics{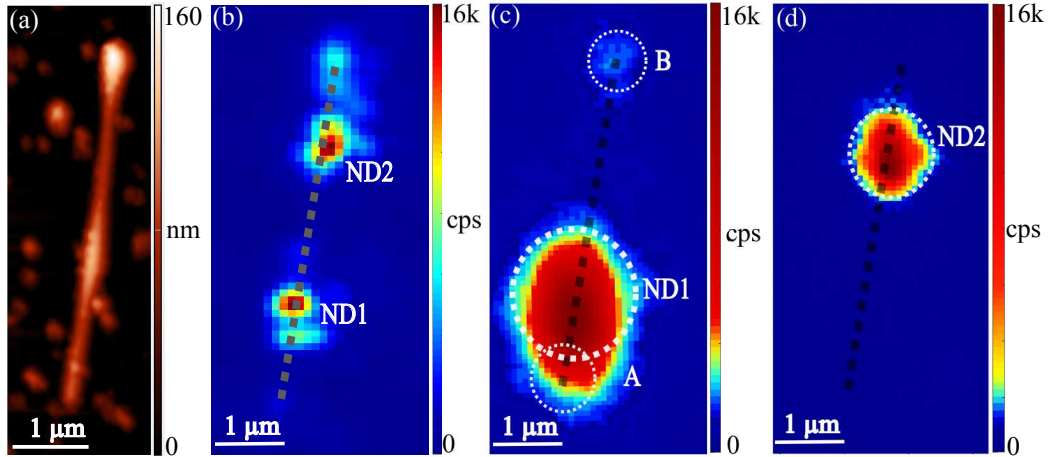}
\caption{(a) AFM image of nanodiamonds and a silver nanowire. (b) a confocal fluorescence microscope image of the  area that includes area in (a). (c) A galvanometric mirror scan image while ND1 is excited with a CW laser. Spot corresponding to ND1 is elongated due to emission from the lower end of the nanowire, indicated as A. Emission from the other end of the nanowire can also be seen clearly, and is indicated as B. (d) A galvanometric mirror scan image while ND2 is excited with a CW laser. The dashed lines in (b) (c) and (d) indicate the position of the silver nanowire. \label{figure3}}
\end{figure}

To couple SiV centers in a nanodiamond to the plasmonic mode of a silver nanowire, we mixed the nanowire solution and the nanodiamond solution, and ultrasonicated the mixture for 0.5 hour before spin coating the mixture on a fused silica glass substrate. This way, we have been able to obtain the SiV-silver nanowire coupled systems. In Fig.~\ref{figure3}(a), we present an AFM image of a silver nanowire, diameter estimated 88~nm as the minimum height of the wire, which is surrounded by nanodiamonds. Some of the nanodiamonds contain SiV centers and appear as bright spots in the confocal fluorescence scan, as presented in Fig.~\ref{figure3}(b). ND1 and ND2 were both found to have a spectrum of SiV center. We, then, excited the nanodiamonds (ND1 and ND2) and took a fluorescence map of the focal plane using galvanometric mirrors. When we excited ND1, we could observe the emission from the far end of the nanowire. The emission from distal end suggests that SiV centers in the nanodiamond excite the silver nanowire mode, plasmons propagate to the wire end and get scattered. In addition, the spot corresponding to ND1 is observed to be elongated. This we infer, is due to the emission from the lower end of the nanowire. The fluorescence map is shown in Fig.~\ref{figure3}(c). However, when ND2 was excited, we did not observe emission from the ends of the nanowire, as presented in Fig.~\ref{figure3}(d). This is due to a relatively large distance between the nanodiamond and the silver nanowire (estimated to be $>$100 nm from fluorescence image).

\begin{figure}[htbp]
\includegraphics{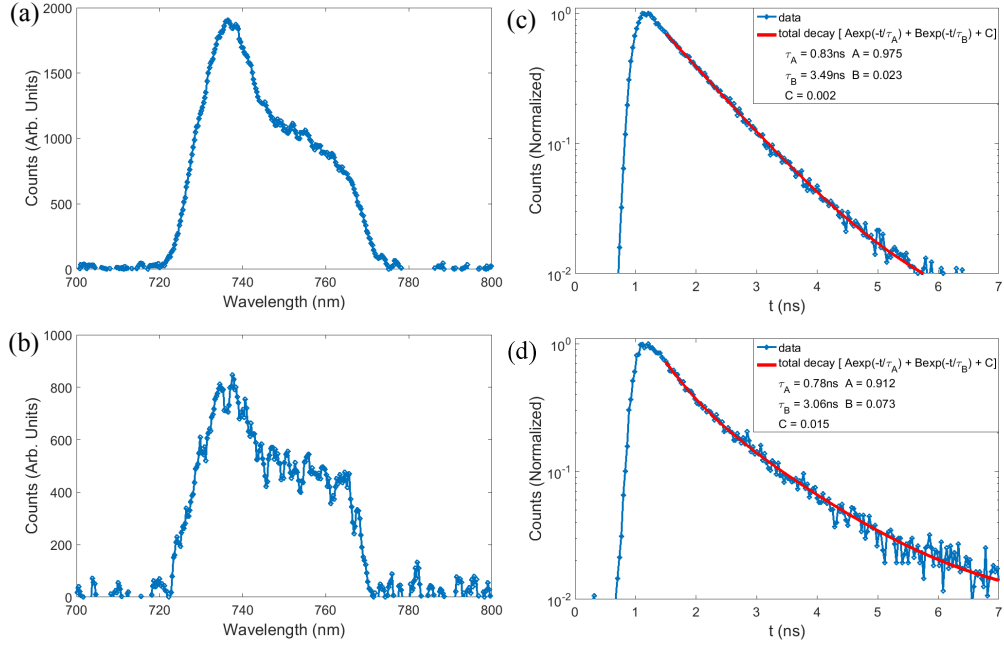}
\caption{(a) and (b) Spectra measured for spots ND1 and B, respectively, in Fig. 3(c). (c) and (d) Lifetime measurement data and a double exponential fit to the data measured for spots ND1 and B, respectively, in Fig. 3(c). \label{figure4}}
\end{figure}

The coupled system was characterized further by measuring the lifetimes and spectrum of different spots in Fig.~\ref{figure3}(c).   The spectra from the two spots in Fig.~\ref{figure3}(c) are shown in Fig.~\ref{figure4}(a) and ~\ref{figure4}(b). The spectra observed at both the spots are typical of SiV centers at room temperature. Lifetime measurement data and a two exponential fit for both the spots are shown in  Fig.~\ref{figure4}(c) and ~\ref{figure4}(d). By comparing the lifetimes from those two spots, one can observe that the lifetime at the distal end (end B) is slightly smaller than that at the ND spot. This is due to higher contribution from the SiV centers which are closer to the silver nanowire, are better coupled and hence have smaller lifetime. 

We have characterized five SiV center-silver nanowire coupled systems in a similar manner. One such coupled system, where nanodiamond containing SiV centers was situated in the middle of the nanowire, is presented in Appendix. There, it can again be observed that the decay lifetime is smaller for the SiV center in the coupled system, when compared to uncoupled SiV centers. Also, when the nanodiamond is excited, emission can be observed from the two distal ends. The shortened lifetime and emission from distal ends confirm the coupling of SiV centers to Silver nanowire modes.The lifetime observed for the coupled SiV centers, averaged over 5 coupled systems, are $\sim$1.5 times smaller than that for the uncoupled SiV centers.

\section{Numerical simulation and discussion}
\begin{figure}[htbp]
\centering\includegraphics{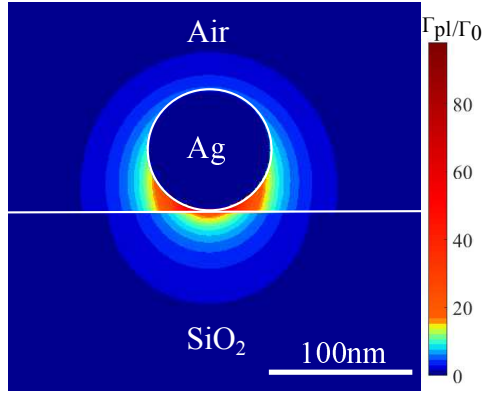}
\caption{Decay rate into the plasmonic mode normalized to the decay rate of the emitter in vacuum.\label{figure5}}
\end{figure}

We obtain the expected decay rate into the plasmonic mode for a silver nanowire, with a diameter 88~nm, placed on a silica substrate by using a numerical model. Plasmonic decay rate,  $\Gamma_{pl},$ relative to the rate in vacuum, $\Gamma_{0},$ is given by the equation,

\begin{equation}
\frac{\Gamma_{pl}}{\Gamma_0} = \frac{3 \pi c \epsilon_0 \left|\mathbf{E}\left(x,y\right) \cdot  \mathbf{n}_D\right|^2}{Re\left\{k_0^2\int_{A_{\infty}} \left( \mathbf{E}\times \mathbf{H}^{\ast}\right) \cdot \mathbf{z}_0 dA\right\}} 
\label{GaPl}
\end{equation}
where c is velocity of light in vacuum, $\epsilon_0$ is the vacuum permittivity, $\mathbf{E}$ and $\mathbf{H}$ are the electric and magnetic fields of the plasmon mode, respectively. $\mathbf{n}_D$ is a unit vector along the dipole emitter, $k_0 = 2\pi/\lambda_0$ is the wavenumber in vacuum, $\mathbf{z}_0$ is a unit vector along the propagation direction (z), $Re$ denotes the real part, and $A_{\infty}$ denotes integration over the transverse plane (xy) ~\cite{2010Chen, 2010Superrad}. The electric and magnetic fields are calculated using a commercial software (COMSOL Multiphysics). For silica n=1.46, and for silver n = 0.147, k = 4.799 ~\cite{Palik}, where n and k are the real and imaginary parts of refractive indices, are used. In Fig.~\ref{figure5}, we show the decay rate of the emitter to the plasmons normalized to the rate in vacuum, maximized over dipole orientation. We note that the nanodiamond structure is not considered in the simulations, because their arbitrary shape makes it a complex problem. The presence of nanodiamond in the vicinity of silver wire can change the decay rates obtained. Nonetheless, as shown in ref.~\cite{2015Esteban}, the presence of the ND shell does not affect considerably the results obtained in the case of a bare dipole source.

From Fig. \ref{figure5}, one can observe that the expected coupling of SiV centers is similar to that of NV centers for similar diameters~\cite{2009Kolesov, 2011Huck, 2013ShkuNL}. However, the measured change in lifetime is smaller than that observed for NV centers, or quantum dots coupled to single silver nanowires\cite{2007Akimov, 2009Kolesov, 2011Huck}. This could be due to lower quantum efficiency of these emitters~\cite{NcommFedor, 2013Elke}. The emitter can decay radiatively with a rate $\Gamma_{r}$ and non radiatively with a rate $\Gamma_{nr}$, when not coupled to a plasmonic waveguide. So, the intrinsic quantum efficiency for the emitter is given by $\eta =\Gamma_{r}/(\Gamma_{r}+\Gamma_{nr})$, and the lifetime is $\tau = 1/(\Gamma_{r}+\Gamma_{nr}).$ When the emitter is coupled to a plasmonic waveguide, the total decay rate is modified and there is one additional channel of decay, decay into plasmons $\Gamma_{pl},$ as presented before. The decay into the radiative modes can be modified by the metallic wire and non radiative decay channels due to the proximity to metal can also appear. So, the modified lifetme $\tau^m =1/(\Gamma^m_{r}+\Gamma_{pl} + \Gamma^m_{nr} + \Gamma_{nr})$, where $\Gamma^m_{r}$ is modified radiative rate of the emitter, and $\Gamma^m_{nr}$ is nonradiative decay rate due to the proximity of emitter to the plasmonic waveguide. For an emitter-nanowire distance  $>5$~nm, $\Gamma_{pl} > \Gamma^m_{nr}$ and $\Gamma^m_{r} \approx \Gamma_{r}$~\cite{2007PRBChang}, so $\tau^m \approx 1/(\Gamma_{r}+\Gamma_{pl} + \Gamma_{nr})$. Therefore, the lifetime ratio of emitter before and after coupling to the plasmonic mode of a nanowire can be written as $\tau/\tau^m \approx 1+\eta(\Gamma_{pl}/\Gamma_{r})$, and the quantum efficiency of emitter coupled to plasmonic waveguide, $\eta_{pl}  \approx (\Gamma_{r}+\Gamma_{pl})/(\Gamma_{r}+\Gamma_{pl}+\Gamma_{nr})$, is higher than that for the uncoupled emitter. As an example, if we consider an optimistic value of $10\%$ for the intrinsic quantum efficiency of SiV centers~\cite{2013Elke}, then $\Gamma_{pl} =5\Gamma_{r}$ on average for our coupled system, which is comparable to the change in lifetimes that has been observed for NV-centers in nanodiamonds~\cite{2009Kolesov, 2011Huck}. 

\section{Conclusion}

We have observed the coupling of SiV centers to silver nanowires. A change in lifetime for the coupled SiV centers was observed. These changes were, however, smaller than those observed for NV centers or quantum dots coupled to a single silver nanowire. This could be due to lower quantum efficiency of SiV centers, as has been pointed out before~\cite{NcommFedor}. In fact, the coupling of SiV centers to silver nanowires increases their quantum efficiency. Even though we have not observed coupling of single SiV centers, it can still be useful for the observation of nonlinearity at single photon level as well as entanglement of two SiV centers coupled to a plasmonic waveguide, as has been observed for five SiV centers coupled to a diamond nanophotonic waveguide. This is possible because at low temperatures transition corresponding to different SiV centers can be separated~\cite{Sipahigil847}. Almost lifetime limited linewidths for optical transitions have been observed at low temperature (below 8K) for SiV centers in nanodiamonds, which were fabricated the same way as the nanodiamonds used in our experiments~\cite{2016Uwe}.  As outlook, SiV centers coupled to plasmonic modes can be used to study nonlinear effects at single photon level and entanglement of SiV centers mediated through plasmonic waveguides. Furthermore, the quantum efficiency of SiV centers might be improved by a change in their chemical environment. SiV centers can be contained in nanodiamonds of sizes around 2 nm, which can make the placement of  SiV centers more precise~\cite{2014Naturenano}. Defect centers in smaller diamonds will also allow for their coupling to gap plasmon modes, which will result in more efficient coupling~\cite{2013ShkuNL}.

\section*{Appendix}
\subsection*{A1. Spectra obtained for SiV centers in nanodiamonds}

\begin{figure}[htbp]
\includegraphics{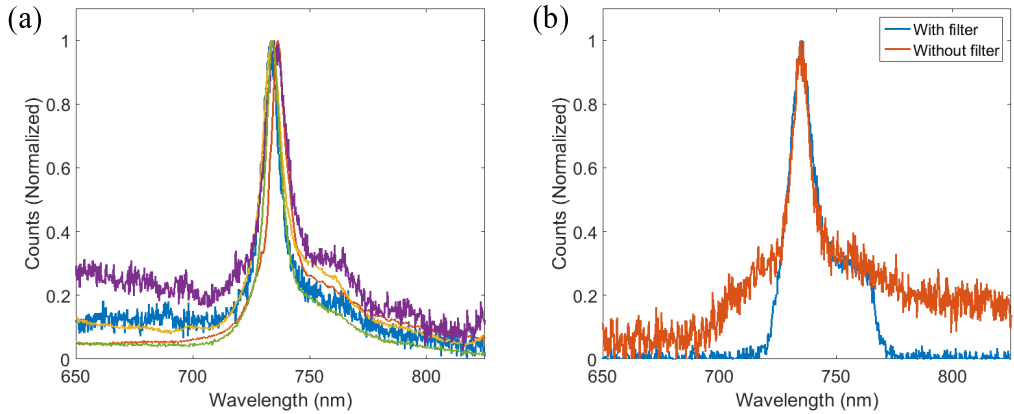}
\caption{(a) Spectra measured for five different nanodiamonds containing SiV centers. (b) Spectra obtained with filters allowing light in the wavelength range between 725~nm and 770~nm and without those filters. \label{figure6}}
\end{figure}

\begin{figure}[htbp]
\includegraphics{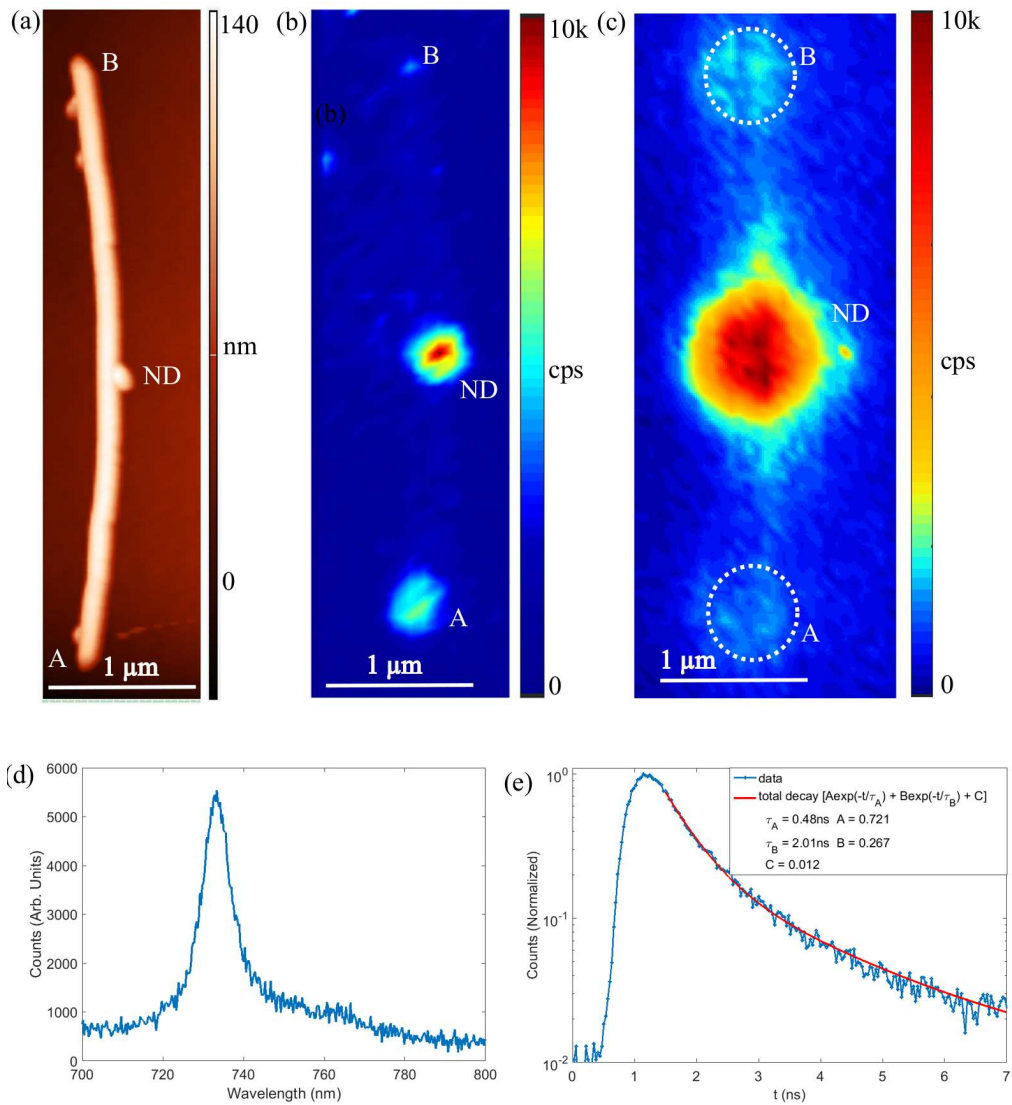}
\caption{(a) AFM image of a SiV center silver nanowire coupled system. Nanodiamond (ND) and the two ends are indicated in the figure. (b) Confocal fluorescence image of the coupled system. (c) A galvanometric mirror scan image while ND is excited with the CW laser at 532~nm. (d) Spectrum measured for spot  ND in (c). (e)  Lifetime measurement data and a double exponential fit to the data measured for spot ND in (c). \label{figure7}}
\end{figure}

The spectrum obtained for different SiV centers can be slightly different, as has been observed before in ref.~\cite{2014Naturenano, 2010Elke}. In Fig.~\ref{figure6}(a), we present normalized spectra measured for five different nanodiamonds. It is clear that the position of peak changes by a few nanometers around 737 nm. Also the shape of the spectra can be different as can be seen clearly and has been observed in ref.~\cite{2014Naturenano, 2010Elke}.

For the measurements presented in the main text, filters allowing light in the wavelength range 725~nm and 770~nm were used. In figure ~\ref{figure6}(b), we can see that the filters allow the SiV center emission and blocks the noise that can be there due to the background emission, which can be present due to other defects in diamond.

\subsection*{A2. Another silver nanowire-SiV centers in a nanodiamond system}

In this section, we present another silver nanowire-SiV centers in a nanodiamond system. An AFM image of the system can be seen in Fig.~\ref{figure7}(a). In the confocal fluorescence image, Fig.~\ref{figure7}(b), the emission from nanodiamond (ND) and one of the wire ends can be seen  clearly. When the ND is excited and a galvanometric mirror scan image is obtained, emission from both the wire ends can be observed, Fig.~\ref{figure7}(b), which clearly suggests that SiV centers in the nanodiamond excite the silver nanowire mode, plasmons propagate to the wire ends and get scattered. The ND has a typical spectrum of SiV centers and the lifetime for the coupled ND is 0.48~ns. A longer lifetime component can also be observed which we think could be due to other defects as a filter for the wavelength range between 725~nm and 770~nm was not used for these measurements, a long pass filter for 700~nm was used instead. Lifetime and spectrum from the distal ends could not be measured for this system as the SiV centers bleached out after the measurement of lifetime at spot ND. The blinking and bleaching of SiV centers have been observed for nanodiamonds produced in the same way~\cite{2016Uwe}.

\section*{Funding}

European Research Council, Grant No. 341054 (PLAQNAP); Russian Foundation for Basic Research, Grant No. 15-03-04490.

\end{document}